\documentclass{PoS}

\title{THE GAMMA-400 SPACE OBSERVATORY: STATUS AND PERSPECTIVES}

\ShortTitle{The GAMMA-400 experiment}

\author{\speaker{O. Adriani, on behalf of the GAMMA-400 collaboration}\\
        University of Florence and INFN Sezione di Firenze\\
        E-mail: \email{adriani@fi.infn.it}}

\author{
A.M. Galper$^{a,b}$, V. Bonvicini$^{c}$ N.P. Topchiev$^{a}$ O. Adriani$^{d}$ R.L. Aptekar$^{e}$ I.V. Arkhangelskaja$^{b}$ A.I. Arkhangelskiy$^{b}$ L. Bergstrom$^{f}$ E. Berti$^{d}$ G. Bigongiari$^{g}$ S.G. Bobkov$^{h}$ M. Boezio$^{c}$ E.A. Bogomolov$^{e}$ S. Bonechi$^{g}$ M. Bongi$^{d}$ S. Bottai$^{d}$ K.A. Boyarchuk$^{i}$ G. Castellini$^{j}$ P.W. Cattaneo$^{k}$ P. Cumani$^{c}$ G.L. Dedenko$^{b}$ C. De Donato$^{l}$ V.A. Dogiel$^{a}$ M.S. Gorbunov$^{h}$ Yu.V. Gusakov$^{a}$ B.I. Hnatyk$^{n}$ V.V. Kadilin$^{b}$ V.A. Kaplin$^{b}$ A.A. Kaplun$^{b}$ M.D. Kheymits$^{b}$ V.E. Korepanov$^{o}$ J. Larsson$^{m}$ A.A. Leonov$^{b}$ V.A. Loginov$^{b}$ F. Longo$^{c}$ P. Maestro$^{g}$ P.S. Marrocchesi$^{g}$ V.V. Mikhailov$^{b}$ E. Mocchiutti$^{c}$ A.A. Moiseev$^{p}$ N. Mori$^{d}$ I.V. Moskalenko$^{q}$ P.Yu. Naumov$^{b}$ P. Papini$^{d}$ M. Pearce$^{m}$ P. Picozza$^{l}$ A.V. Popov$^{h}$ A. Rappoldi$^{k}$ S. Ricciarini$^{j}$ M.F. Runtso$^{b}$ F. Ryde$^{m}$ O.V. Serdin$^{h}$ R. Sparvoli$^{l}$ P. Spillantini$^{d}$ S.I. Suchkov$^{a}$ M. Tavani$^{r}$ A.A. Taraskin$^{b}$ A. Tiberio$^{d}$ E.M. Tyurin$^{b}$ M.V. Ulanov$^{e}$ A. Vacchi$^{c}$ E. Vannuccini$^{d}$ G.I. Vasilyev$^{e}$ Yu.T. Yurkin$^{b}$ N. Zampa$^{c}$ V.N. Zirakashvili$^{s}$ and V.G. Zverev$^{b}$
\\
\llap{$^{a}$} Lebedev Physical Institute, Russian Academy of Sciences, Moscow, Russia\\
\llap{$^{b}$} National Research Nuclear University MEPhI, Moscow, Russia\\
\llap{$^{c}$} INFN, Sezione di Trieste and Physics Department of University of Trieste, Trieste, Italy\\
\llap{$^{d}$} INFN, Sezione di Firenze and Physics Department of University of Florence, Firenze, Italy\\
\llap{$^{e}$} Ioffe Institute, Russian Academy of Sciences, St. Petersburg, Russia\\
\llap{$^{f}$} Stockholm University, Department of Physics; and the Oskar Klein Centre, AlbaNova University Center, Stockholm, Sweden\\
\llap{$^{g}$} Department of Physical Sciences, Earth and Environment, University of Siena and INFN, Sezione di Pisa, Italy\\
\llap{$^{h}$} Scientific Research Institute for System Analysis, Russian Academy of Sciences, Moscow, Russia\\
\llap{$^{i}$} Research Institute for Electromechanics, Istra, Moscow region, Russia\\
\llap{$^{j}$} IFAC- CNR and Istituto Nazionale di Fisica Nucleare, Sezione di Firenze, Firenze, Italy\\
\llap{$^{k}$} INFN, Sezione di Pavia, Pavia, Italy\\
\llap{$^{l}$} INFN, Sezione di Roma 2 and Physics Department of University of Rome Tor Vergata, Italy\\
\llap{$^{m}$} KTH Royal Institute of Technology, Department of Physics; and the Oskar Klein Centre, AlbaNova University Center, Stockholm, Sweden\\
\llap{$^{n}$} Taras Shevchenko National University of Kyiv, Kyiv, Ukraine\\
\llap{$^{o}$} Lviv Center of Institute of Space Research, Lviv, Ukraine\\
\llap{$^{p}$} CRESST/GSFC and University of Maryland, College Park, Maryland, USA\\
\llap{$^{q}$} Hansen Experimental Physics Laboratory and Kavli Institute for Particle Astrophysics and Cosmology, Stanford University, Stanford, USA\\
\llap{$^{r}$} Istituto Nazionale di Astrofisica IASF and Physics Department of University of Rome Tor Vergata, Rome, Italy\\
\llap{$^{s}$} Pushkov Institute of Terrestrial Magnetism, Ionosphere, and Radiowave Propagation, Troitsk, Moscow region, Russia\\ 
}

\abstract{The present design of the new space observatory  GAMMA-400 is presented in this paper. The instrument has been designed for the optimal detection of gamma rays in a broad energy range (from $\sim$100 MeV up to 3 TeV), with excellent angular and energy resolution. The observatory will also allow precise and high statistic studies of the electron component in the cosmic rays up to the multi TeV region, as well as protons and nuclei spectra up to the knee region. The GAMMA-400 observatory will allow to address a broad range of science topics, like search for signatures of dark matter, studies of Galactic and extragalactic gamma-ray sources, Galactic and extragalactic
diffuse emission, gamma-ray bursts and charged cosmic rays acceleration and
diffusion mechanism up to the knee.}

\FullConference{Science with the New Generation of High Energy Gamma-ray experiments, 10th Workshop - Scineghe2014\\
		04-06 June 2014\\
		Lisbon - Portugal}

\begin{document}

\section{The GAMMA-400 mission}
The GAMMA-400 space mission has been approved by the Russian Space Agency to deeply and precisely investigating many important aspects in the gamma ray astronomy, cosmic ray science and high energy charged particle spectra measurement.

The GAMMA-400 experiment will be installed onboard of the 'Navigator' space platform, manufactured by the NPO Lavochkin Association, able to accommodate high mass - large volume scientific payload. The Navigator platform is equipped with all the systems necessary for a standalone space system:
\begin {itemize}
\item Main radio complex;
\item Attitude control system;
\item Power supply system;
\item Thermal control system;
\item Autonomous electronics module;
\item Low gain telemetry antenna and feed system;
\item High data rate communication radio link;
\item Up-down phase transfer radio link;
\item Solar panel attitude control system;
\item Orbit correction engines.
\end{itemize}
The launch date currently foreseen is at the beginning of 2020s. 

The use of the Navigator spacecraft will make the GAMMA-400 a really unique opportunity for the near future cosmic ray science, since it allows the installation of a very large scientific payload (mass $\sim$4100 kg, available electric power $\sim$2000 W, downlink telemetry $\sim$100 GB/day, lifetime > 7 years), able to significantly contribute to the next generation instrument for the gamma-ray astronomy and cosmic ray physics. 

The GAMMA-400 experiment will be initially installed on a highly elliptical orbit (with apogee 300.000 km and perigee 500 km, with an inclination of 51.4$^\circ$), with 7 days orbital period; the orbit will then be gradually changed, in 5 months, to a circular one, with $\sim$200.000 km radius. A great advantage of such an orbit is the fact that the full sky coverage will always be available for gamma ray astronomy, since the Earth will not cover a significant fraction of the sky, as is usually the case for low Earth orbit. The GAMMA-400 source pointing strategy will hence be properly defined to maximize the physics outcome of the experiment. 

\section{The GAMMA-400 apparatus}
The GAMMA-400 apparatus~\cite{Gamma400-1, Gamma400-2} on board of the Navigator platform will be composed by:
\begin{itemize}
\item the main cosmic ray telescope, described in detail in the next paragraph;
\item two star sensors, with 5 arc-second accuracy;
\item the 'Konus-FG' Gamma-Ray Burst monitor systems, under the responsibility of the St. Petersburg Ioffe Physical Technical Institute;
\item four direction detectors, located on telescopic booms;
\item two spectrometric detectors;
\item two magnetometers, produced by the Ukraine Lviv Center of Institute of Space Research.
\end{itemize}

\subsection{The main gamma ray telescope}
The original Russian design for the GAMMA-400 mission was focused on:
\begin{itemize} 
\item Gamma-rays, in the 100 MeV - 3 TeV energy range
\item High energy electrons+positrons up to the TeV region
\end{itemize}
with the aim of
\begin{itemize}
\item  studying the nature and features of weakly interacting massive particles, that could compose the Dark Matter
\item studying the nature and the features of variable gamma-ray activity of astrophysical objects, from stars to galactic clusters
\item studying the mechanisms of generation, acceleration, propagation and interaction of cosmic rays in galactic and intergalactic spaces
\end{itemize}

\begin{figure}
\centering
\includegraphics[width=.6\textwidth]{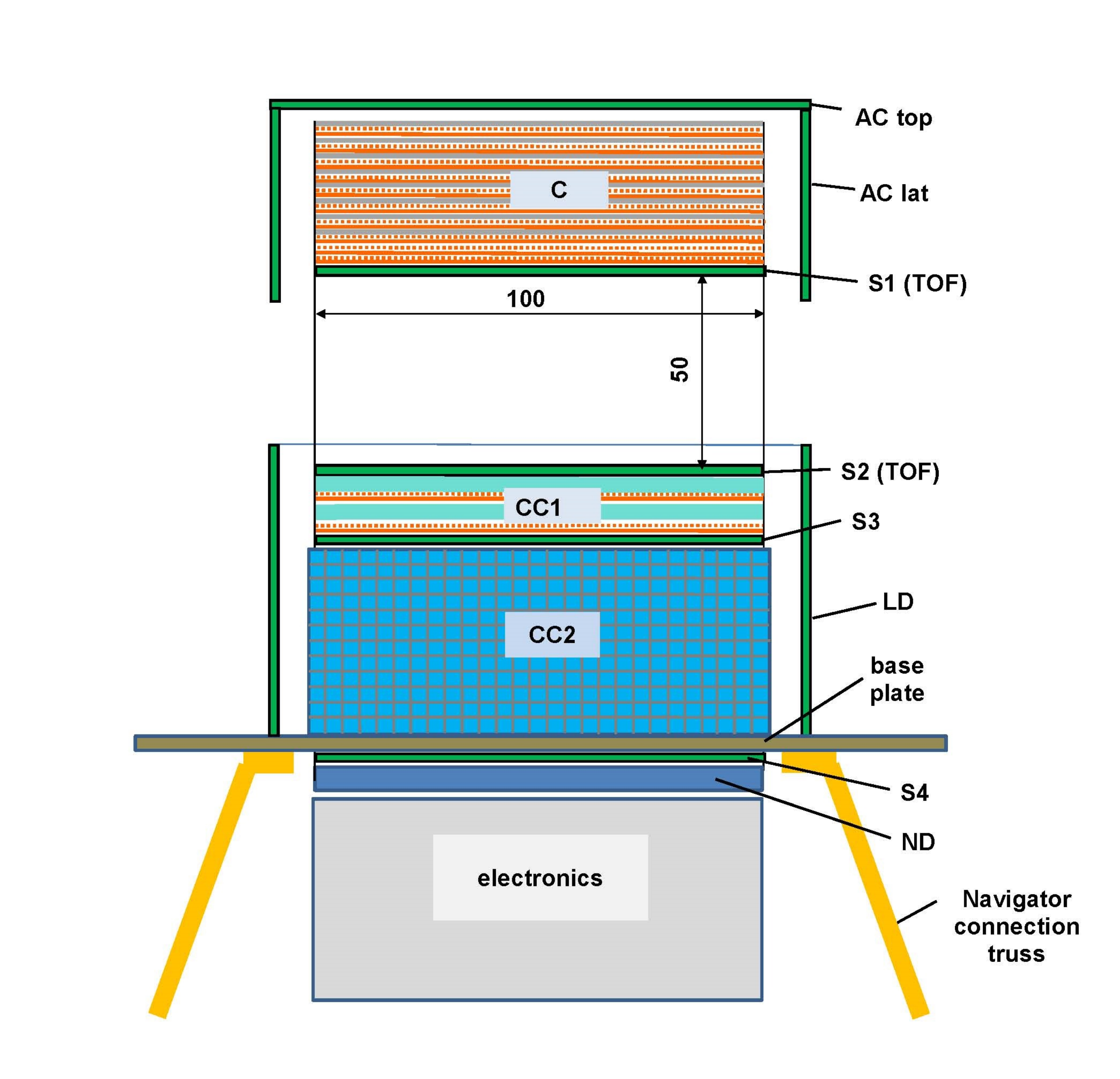}
\caption{Schematic drawing of the GAMMA-400 telescope.}
\label{Fig1}
\end{figure}

However, during the last years, the collaboration between Italian and Russian groups has resulted in a new version of the apparatus, to significantly improve the GAMMA-400 design and performance. The basic idea was to develop a jointly defined dual instrument that, taking into account the currently available financial resources, optimizes the scientific performance and improves them with respect to the original version. 
Figure~\ref{Fig1} schematically describe the current version of the GAMMA-400 telescope.

\begin{figure}
\centering
\includegraphics[width=.6\textwidth]{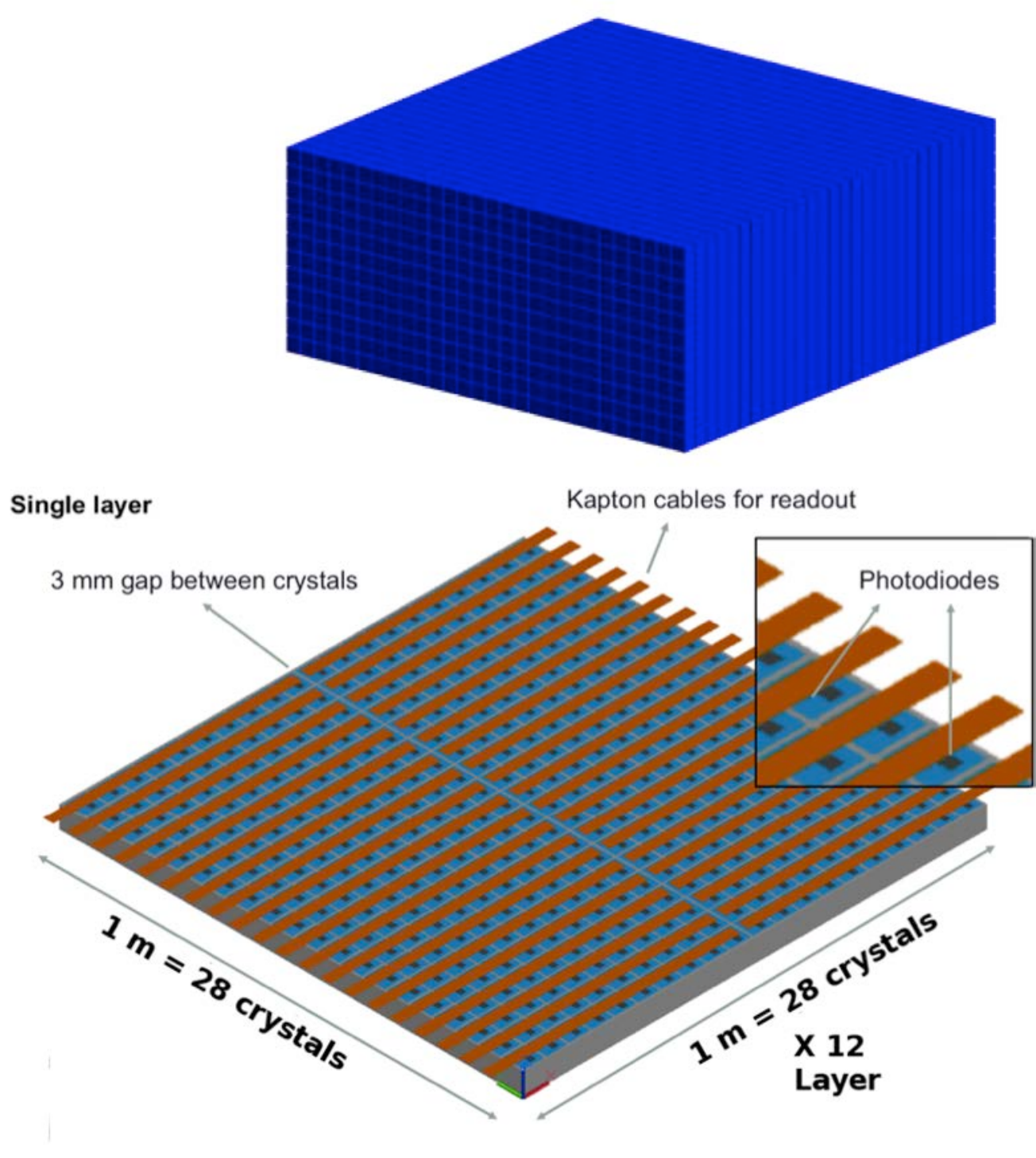}
\caption{Schematic drawing with some details of the CC2 calorimeter.}
\label{Fig2}
\end{figure}

Starting from the top, the telescope is composed by:
\begin {itemize}
\item an anticoincidence system (AC), composed by plastic scintillators read out with Silicon Photomultipliers, located both on top and on the lateral side of the apparatus. The system is essentially used to discriminate between charged and neutral particles.
\item a converter-tracker system (C), realized by means of 10 double layers of microstrip silicon detectors, interleaved with thin tungsten layers. The system is used to pair convert the photons and to precisely reconstruct the photon direction by the detection in the silicon layers of the $e^+e^-$ pair. In more detail, 8 layers are each made by 0.08 X$_0$ thick tungsten absorber and 2 layers of x-y silicon microstrip sensors, while 2 layers are made only by active material, without any tungsten. 
The converter-tracker system is assembled in 4 towers, each 50$\times$50 cm$^2$. The silicon miscrostrip detectors are arranged in ladders, ~50 cm long, readout at their ends. The implant pitch of the sensors is 80 $\mu$m, while the readout pitch is 240 $\mu$m, to reduce the overall number of electronic channels, while maintaining a good spatial resolution by exploiting the capacitive charge division between the implanted strips. 
The overall number of silicon sensors is 2000, corresponding to 153.600 readout channels. The expected power consumption on the front end chips is approximately 80 W.
\item a Time of Flight system (TOF), made by plastic scintlllators coupled to Silicon Photomultiplier, used both to generate the trigger for the apparatus and to reject albedo particles by measuring their velocity.
\item a shallow, finely segmented imaging calorimeter (CC1), used to significantly improve the photon angular resolution by precisely measuring the converted $e^+e^-$ pair after a large 50 cm lever arm below the converter-tracker. CC1 is composed by 2 layers of CsI(Tl) crystals, each 1 X$_0$ deep, interleaved with 2 silicon microstrip layers of the same type used for the converter-tracker. 
\item a deep, isotropic and homogeneous calorimeter (CC2, see Fig.~\ref{Fig2}), made by 28$\times$28$\times$12 small cubic CsI(Tl) crystals (with 3.6 cm side), developed on the basis of the CaloCube project R\&D~\cite{CCube}. The overall dimensions of the calorimeter (1 m $\times$ 1 m $\times$ 0.47 m, corresponding to 
54.6 X$_0 \times$ 54.6 X$_0 \times$ 23.4 X$_0$ and 
2.5 $\lambda_I \times$ 2.5 $\lambda_I \times$ 1.1$\lambda_I$) will allow an excellent containment of electromagnetic particles up to very high energies ($\sim$10 TeV).
The overall mass ($\sim$2000 kg), coupled to the large dimensions, will allow optimal detection of high energy hadrons, up to 10$^{15}$ eV, with an effective geometric factor of the order of 4 m$^2$sr, giving the possibility to directly probe on orbit the knee region. This is accomplished by detecting particles not only on the top surface, but also on the lateral side, hence significantly increasing the overall geometrical factor. 
\item a neutron detector (ND), located below the CC2, used to improve the proton/electron ratio.
\end{itemize}

\section{The expected physics performances}

The introduction of a highly segmented homogeneous calorimeter made with CsI(Tl) cubes, with improved energy resolution and extended geometrical factor, coupled to the improvement of the converter-tracker and CC1 structure (in particular with the analogue readout and reduced pitch of the silicon sensors) make GAMMA-400 an excellent dual instrument, optimized both for photons (in the 100 MeV - 1 TeV energy range) and charged cosmic rays (electrons up to 10 TeV and high energy nuclei, p and He spectra up to the knee region,  10$^{14}$ - 10$^{15}$ eV).

\subsection{Performances in the photon detection}
\begin{figure}
\centering
\includegraphics[width=.7\textwidth]{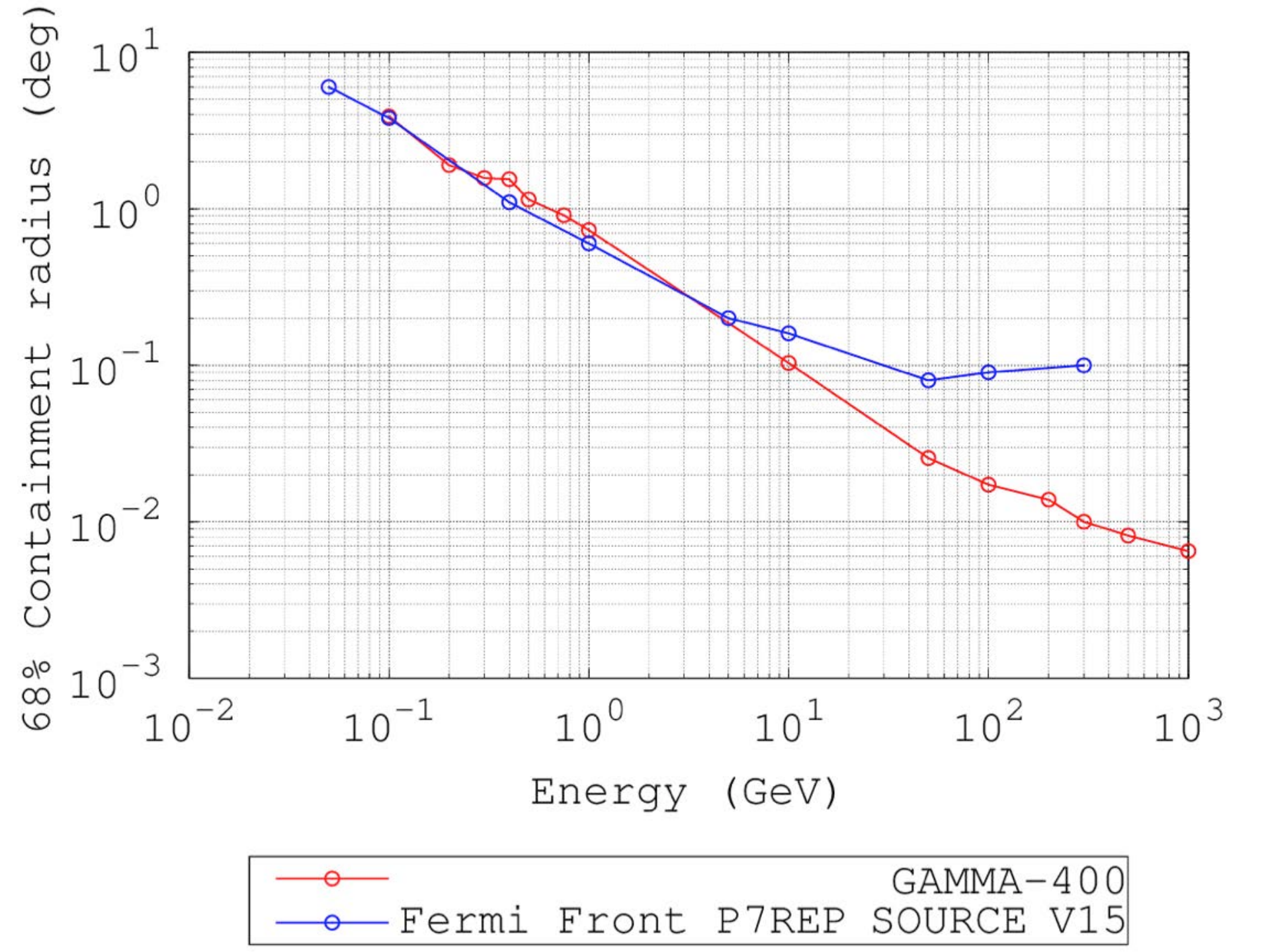}
\caption{Expected angular resolution of GAMMA-400 for photons, compared to Fermi.}
\label{Fig3}
\end{figure}

\begin{figure}
\centering
\includegraphics[width=.7\textwidth]{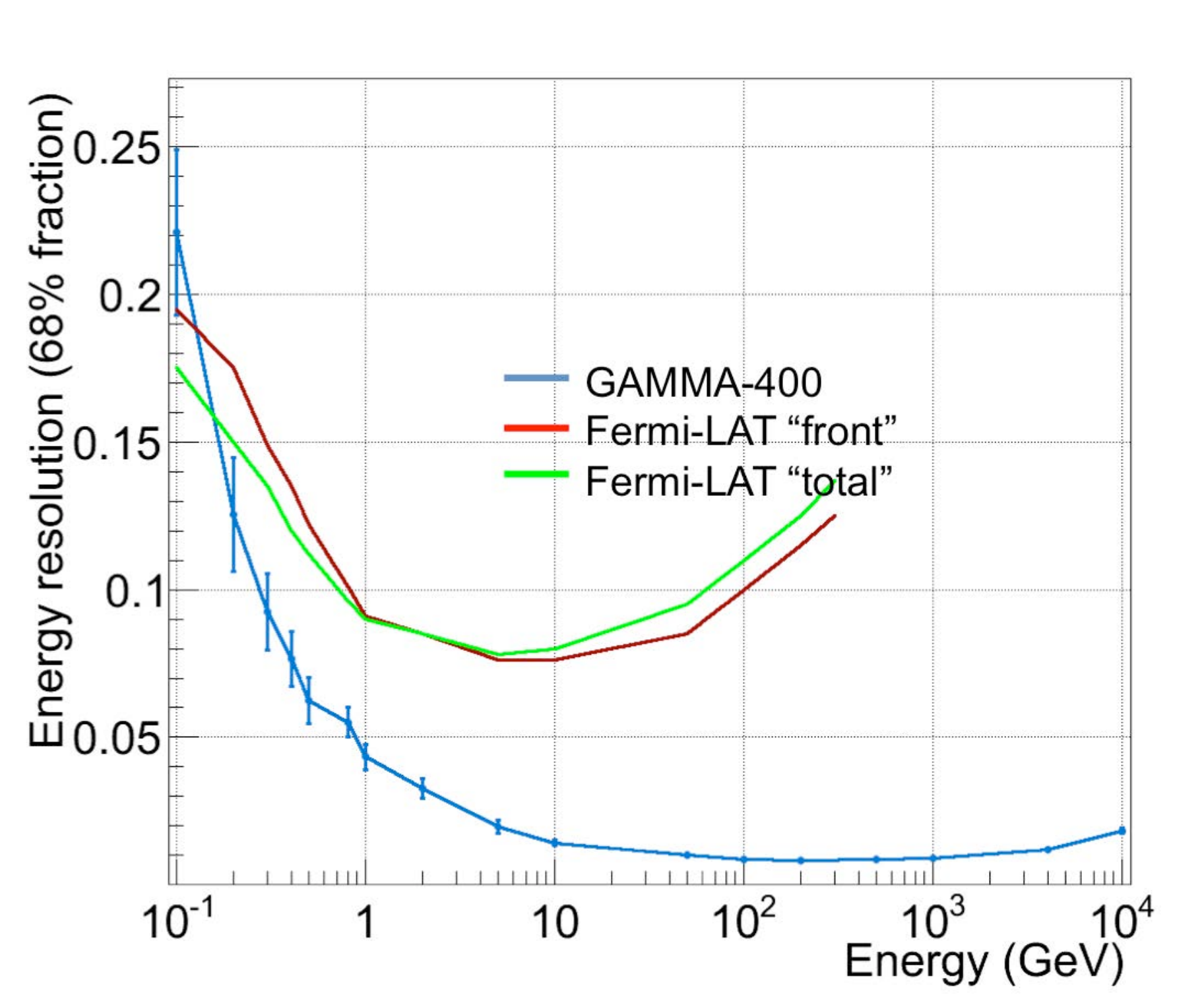}
\caption{Expected energy resolution of GAMMA-400 for photons, compared to Fermi.}
\label{Fig4}
\end{figure}

\begin{figure}
\centering
\includegraphics[width=.7\textwidth]{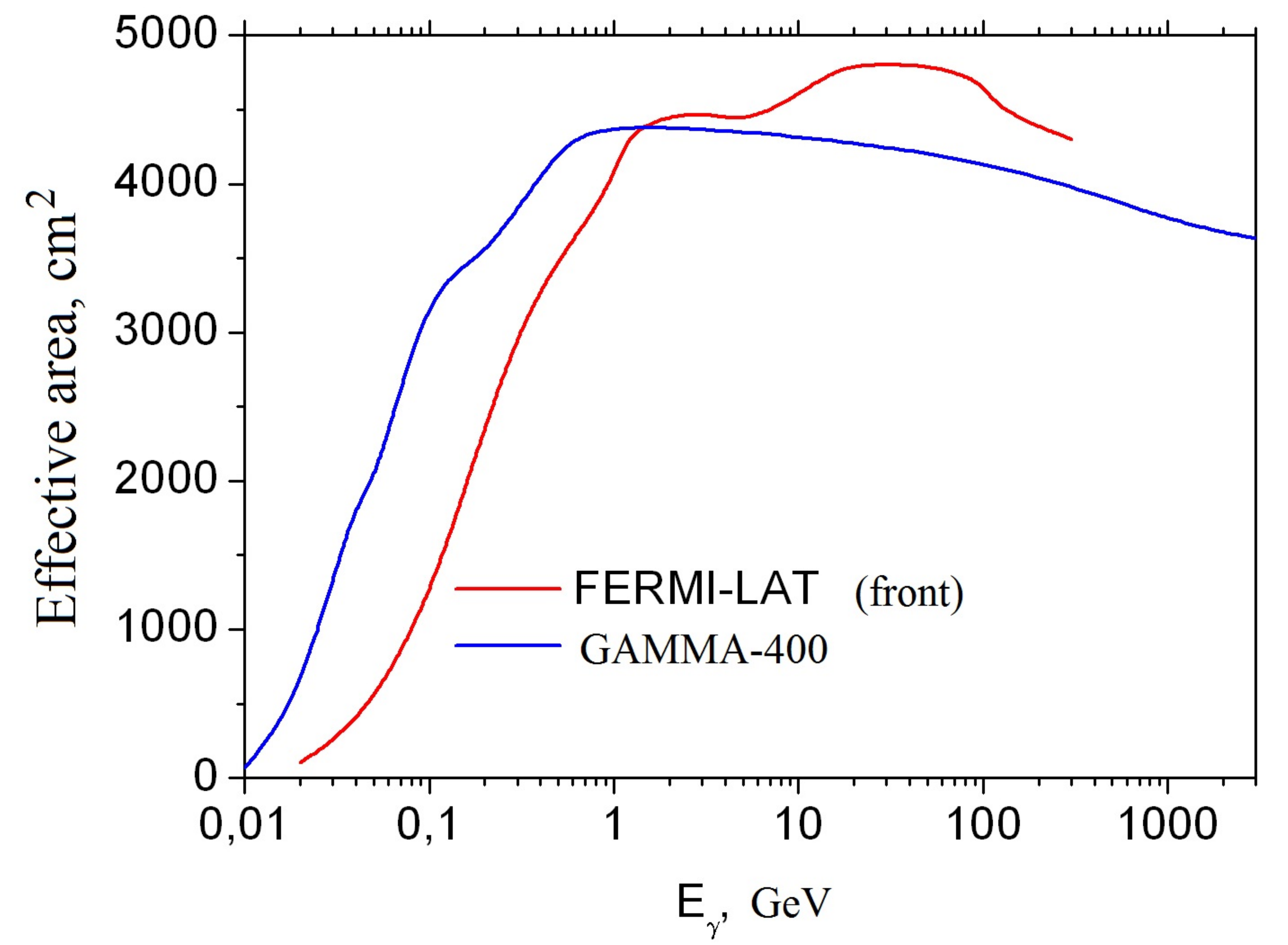}
\caption{Effective area of GAMMA-400 in the photon detection, compared to 'Fermi front'.}
\label{Fig5}
\end{figure}

For what concern the instrument's performances in the gamma detection, Figures~\ref{Fig3},~\ref{Fig4} and~\ref{Fig5} show respectively the angular resolution, the energy resolution and the effective area of the GAMMA-400 instrument, compared with the 'state of the art' Fermi experiment. 

From Figure~\ref{Fig3} we can observe that the angular resolution of GAMMA-400 is comparable to the Fermi one in the 100 MeV - few GeV range, while is significantly better above few GeV. This is accomplished by a proper combination of the excellent spatial resolution and the long lever arm available in between the converter-tracker and the CC1 calorimeter. 

Figure~\ref{Fig4} demonstrate the excellent performances of the instrument for what concern the energy resolution, that clearly overpass the Fermi one above few hundreds MeV, reaching a resolution better than 2\% for the multi GeV range. This is possible thanks to the excellent performances of the homogeneous and very deep calorimeter, able to fully contain electromagnetic shower generated by multi-TeV photons.

The comparison between the effective area of GAMMA-400 and 'Fermi front', reported in Figure~\ref{Fig5}, shows that in the low energy region (below 1 GeV) the effective area of GAMMA-400 is larger than the 'Fermi front' one, while in the high energy region the situation is interchanged. However, for a full comparison of the performances we should take into account the optimized choice of the pointing strategy for important sources or for the whole sky survey, taking into account that the full sky is available to GAMMA-400, due to the peculiar orbit, very far away from any possible earth occultation.
Additionally, the high energy part of the photon spectrum can be investigated with the calorimeter only, with an angular resolution of the order of few degrees. In this way the acceptance can clearly be increased, reaching a level comparable to the 'Fermi total' one ($\sim 8000$ cm$^2$). 

\begin{figure}
\centering
\includegraphics[width=.7\textwidth]{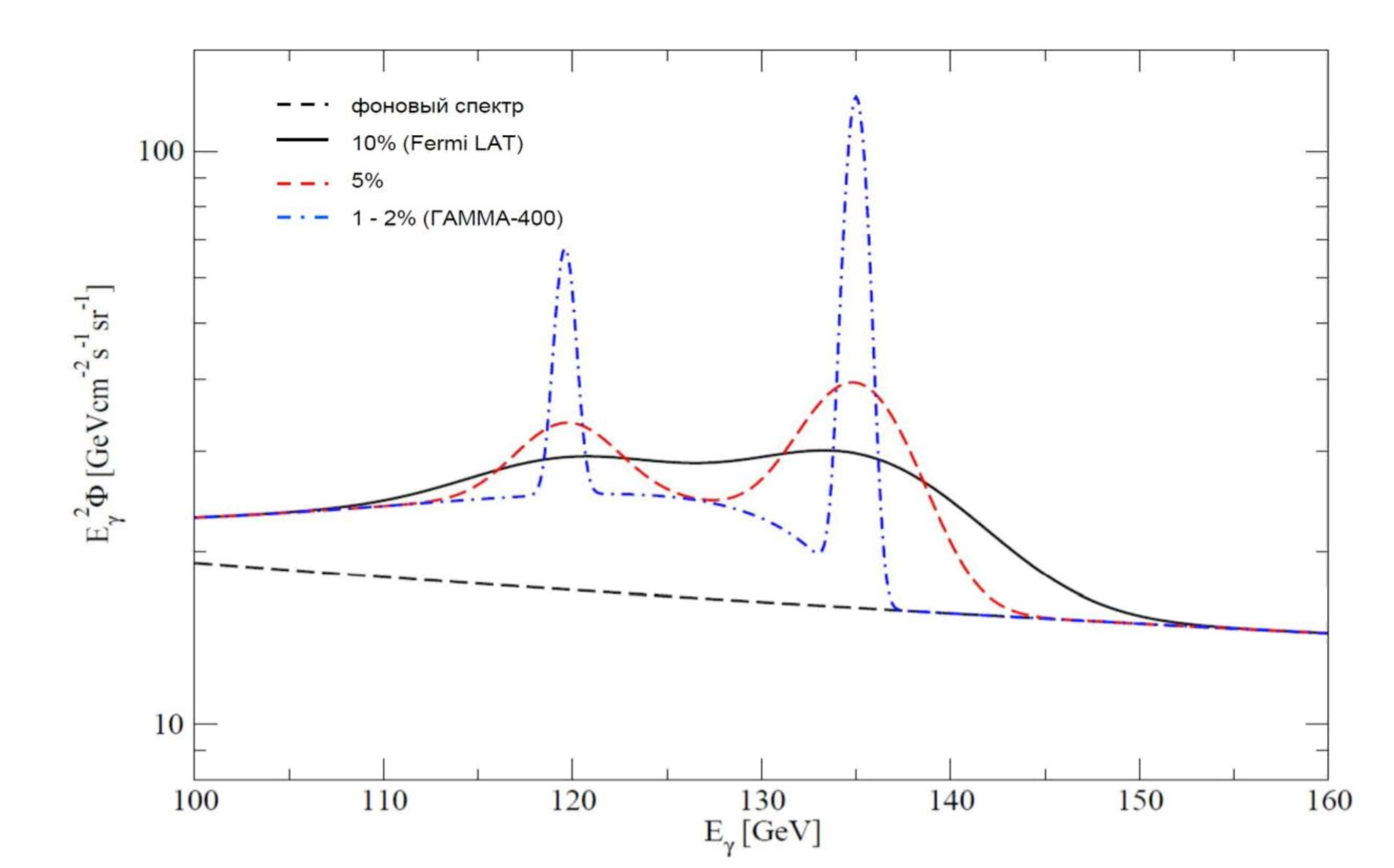}
\caption{
$\gamma$-ray differential energy spectrum expected for a 135 GeV right-handed neutrino dark matter candidate, as can be observed by Fermi LAT and GAMMA-400 (from~\cite{Bergstrom}).}
\label{Fig_berg}
\end{figure}

Thanks to the excellent performances in the photon detection, GAMMA-400 will be able to optimally detect a possible Dark Matter signal, by searching with high resolution for Gamma-ray lines both in the Galactic center region and in the whole sky~\cite{Gamma400-3}. 
Figure~\ref{Fig_berg}~\cite{Bergstrom} shows the $\gamma$-ray differential energy spectrum expected for a possible 135 GeV right-handed neutrino dark matter candidate, as can be observed by Fermi LAT (with 10\% and 5\% energy resolution) and GAMMA-400 (with 1-2\% energy resolution). This figure demonstrates the clear advantage of GAMMA-400 with respect to Fermi, but the real possibility of GAMMA-400 to resolve the 135 GeV line over the background of the diffuse gamma-ray emission requires additional simulation.  

Moreover, the experiment will be able to carry on precise measurements of the high-energy gamma-ray spectrum, that are a clue for a better understanding of the high energy astrophysical phenomena (SuperNova Remnants, Pulsars, Active Galactic Nuclei, Gamma Ray Burst)~\cite{Gamma400-4}.

\subsection{Performances in the charged particles detection}

For what concern the charged particles detection capabilities, the GAMMA-400 instrument has been optimized both in the high energy electron and very high energy hadron detections. 

The full containment electromagnetic calorimeter allow precise measurement of the ($e^+ + e^-$) spectrum up to the multi-TeV region with excellent energy resolution, $\sim$2\%. The high energy part of the spectrum, up to >10 TeV, can however be probed only if primary electron source are existing close to the earth ($\leq$ 1 kpc), due the significant energy loss by radiation. The instrument is hence a really optimized electron detector, giving the possibility to probe the interstellar medium looking to some primary electron source, like, for example, the SuperNova Remnants.

The GAMMA-400 observatory is also an optimal tool for the very high energy hadron detection. For this item two critical parameters exist in the detection system: the acceptance, that should be as large as possible, to have the possibility to measure a significant amount of the $\sim$10$^{15}$ eV particles, and the hadronic energy resolution. The large area homogeneous and isotropic calorimeter, accepting particles from all sides, gives us for the first time the possibility to directly observe in space such high energy hadrons. 

\begin{figure}
\centering
\includegraphics[width=.9\textwidth]{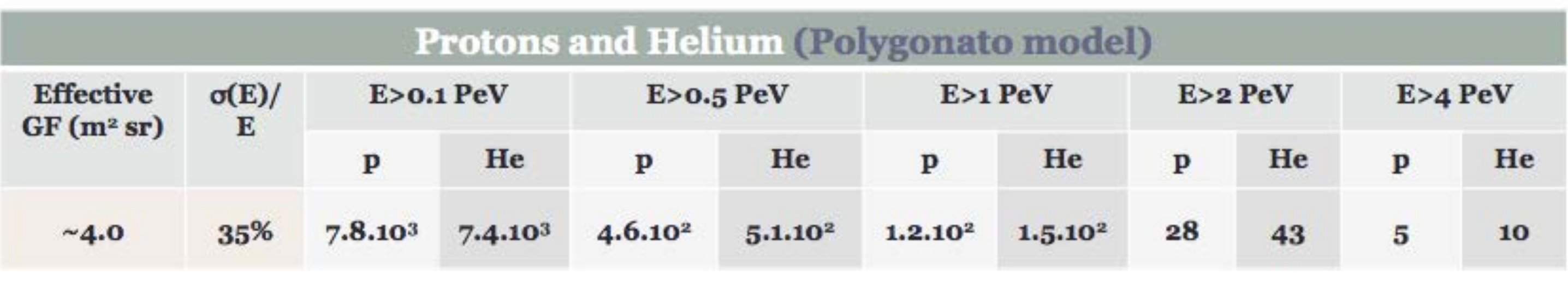}
\caption{Expected number of proton and helium events in 10 years data taking, according to the Polygonato model~\cite{Polygonato}.}
\label{Fig6}
\end{figure}

Figure~\ref{Fig6} shows the expected number of high energy protons and heliums nuclei for a 10 years exposure of the GAMMA-400 experiment, assuming the Polygonato model~\cite{Polygonato}, with an expected p/e rejection factor of 10$^5$, and with a realistic reconstruction efficiency of the order of 40\%. We can observe that we expect to collect a statistic larger than 100 events both for protons and helium, allowing a detailed probe of the interesting knee region, that has never been directly investigated with an in orbit detector up to now. 

\begin{figure}
\centering
\includegraphics[width=.7\textwidth]{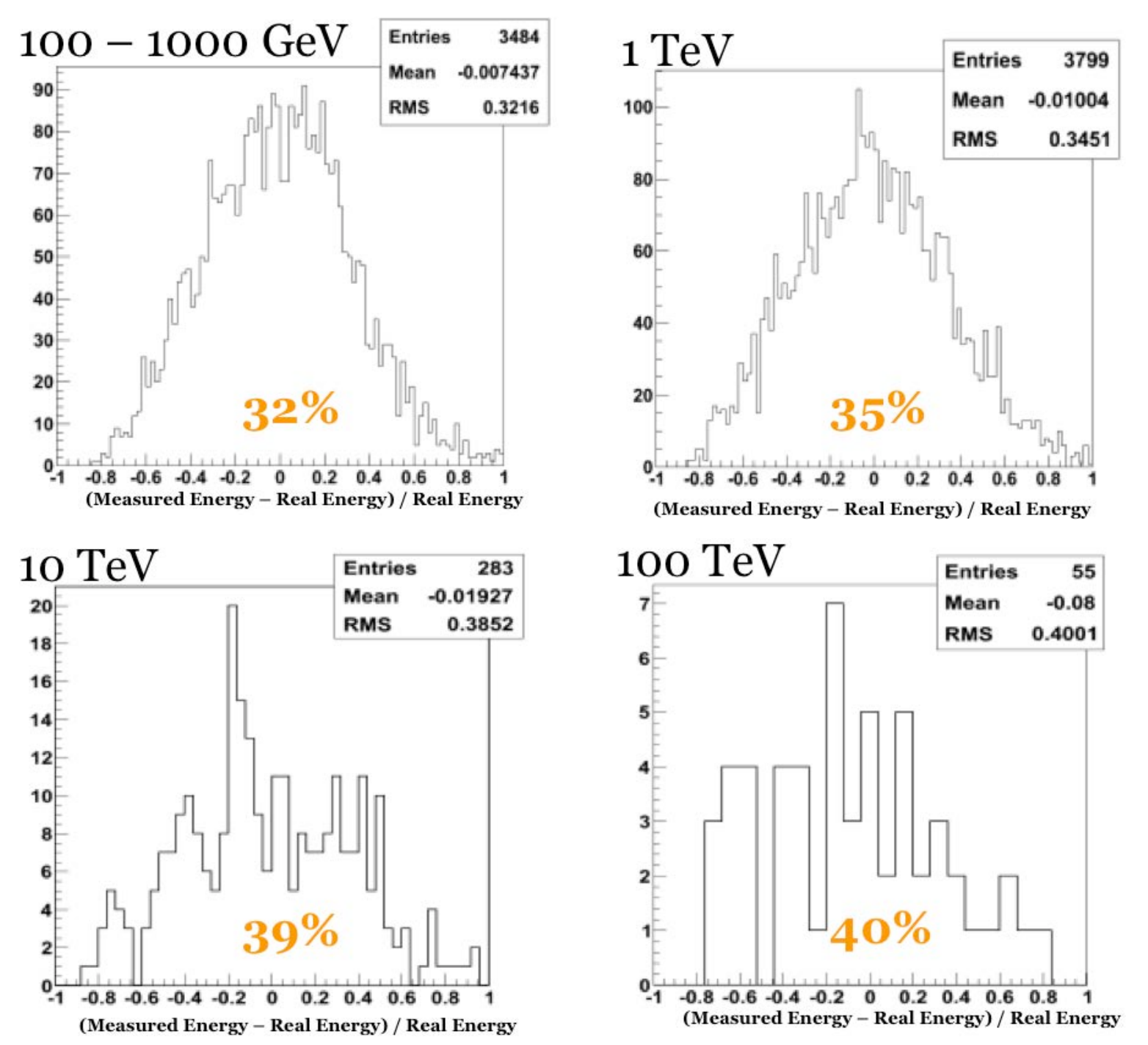}
\caption{Energy resolution for high energy protons.}
\label{Fig7}
\end{figure}

Figure~\ref{Fig7} report the expected energy resolution for high energy protons, from 100 GeV up to 100 TeV. Typical values lies in between 30\% and 40\%, giving us the possibility to study in detail the knee region for the different nuclei. 

In addition to the direct investigation of the knee region, GAMMA-400 can also significantly contribute in the study of the acceleration and propagation mechanism of the cosmic ray in our galaxy, in particular helping to understand the limit of the acceleration phenomena, looking for the sources of the cosmic rays, trying to investigate the dependence of the acceleration and propagation on the charge of the nucleus.

\end{document}